\documentclass[a4paper,fleqn,twocolumn,usenatbib]{mnras}

\usepackage{mathptmx}
\usepackage{ulem}
\usepackage[T1]{fontenc}
\usepackage{ae,aecompl}

\usepackage{graphicx}	
\usepackage{amsmath}	
\usepackage{amssymb}	
\usepackage{indentfirst}

\def\be{\begin{equation}}
\def\ee{\end{equation}}
\def\ba{\begin{eqnarray}}
\def\ea{\end{eqnarray}}

\newcommand\DM{{\rm DM}}
\newcommand\obs{{\rm obs}}
\newcommand\IGM{{\rm IGM}}
\newcommand\MW{{\rm MW}}
\newcommand\e{{\rm e}}

\title[Using FRBs to constrain baryon fraction in IGM]{Reconstruction of Baryon Fraction in Intergalactic Medium through Dispersion Measurements of Fast Radio Bursts}

\author[Dai \& Xia]{
Ji-Ping Dai,$^{1}$
Jun-Qing Xia,$^{1}$\thanks{E-mail: xiajq@bnu.edu.cn}
\\
$^{1}$Department of Astronomy, Beijing Normal University, Beijing, 100875, China
}

\date{Accepted XXX. Received YYY; in original form ZZZ}

\pubyear{2021}

\begin{document}
\label{firstpage}
\pagerange{\pageref{firstpage}--\pageref{lastpage}}
\maketitle

\begin{abstract}
Fast radio bursts (FRBs) probe the total column density of free electrons in the intergalactic medium (IGM) along the path of propagation though the dispersion measures (DMs) which depend on the baryon mass fraction in the IGM, i.e., $f_{\IGM}$. In this paper, we investigate the large-scale clustering information of DMs to study the evolution of $f_{\IGM}$. When combining with the Planck 2018 measurements, we could give tight constraints on the evolution of $f_{\IGM}(z)$ from about $10^4$ FRBs with the intrinsic DM scatter of $30(1+z)~ \rm pc/cm^3$ spanning {80\%} of the sky and redshift range $z=0-3$. Firstly, we consider the Taylor expansion of $f_{\IGM}(z)$ up to second order, and find that the mean relative standard deviation $\sigma(f_{\IGM})\equiv\left\langle \sigma[f_{\IGM}(z)] /f_{\IGM}(z) \right\rangle$ is about {7.2\%}. In order to alleviate the dependence on fiducial model, we also adopt a non-parametric methods in this work, the local principle component analysis. We obtain the consistent, but weaker constraints on the evolution of $f_{\IGM}(z)$, namely the mean relative standard deviation $\sigma(f_{\IGM})$ is {24.2\%}. With the forthcoming surveys, this could be a complimentary method to investigate the baryon mass fraction in the IGM.
\end{abstract}




\section{Introduction}

Until now, the baryon mass fraction in the intergalactic medium is still a poorly known parameter in modern cosmology. Based on the observations, \citet{Fukugita:1997bi} presented an estimation of the global budget of baryons in all states, and they found there are about 17\% of the baryons in the form of stars and their remnants. Afterwards, there are many works trying to use both numerical simulations \citep{Cen:1998hc, Cen:2006by, Meiksin:2007rz} and observations \citep{Fukugita:2004ee, Shull:2011aa, Hill:2016dta} to study the baryon distribution. Unfortunately, $f_{\IGM}$ is still not well understood. \citet{Meiksin:2007rz} showed that about 90\% of the baryons produced by the Big Bang are contained within the IGM at $z\ge1.5$, while \citet{Shull:2011aa} found that $18\pm4 \%$ of the baryons exist in the collapsed phase (galaxies, groups, clusters, circumgalactic medium) at redshift $z\le 0.4$, which means $f_{\IGM}\simeq0.82$. These evidences show $f_{\IGM}$ could be growing with redshift, which is reasonable because there are less massive halos in the early universe \citep{McQuinn:2013tmc}.

Recently, since the first detection of fast ratio burst \citep{Lorimer:2007qn}, it is very interesting to investigate the potential of FRBs in Cosmology \citep{Deng:2013aga, Gao:2014iva, Zhou:2014yta, Fialkov:2017qoz, Li:2017mek}. FRBs are millisecond radio transients characterized by the excess dispersion measure with respect to the Galactic values. Since the DM of FRB is an integration of the total column density of free electrons in the IGM along the line of sight from its source, some previous works simulated the FRBs catalog with precise redshift information to constrain the evolution of baryon fraction in the IGM \citep{Deng:2013aga, Keane:2016yyk, Li:2019klc, Wei:2019uhh}.

On the other hand, there are other papers trying to use the large-scale clustering statistics of dispersion measures of FRBs to study the information of the host environment \citep{Shirasaki:2017otr}, the reionization history \citep{Dai:2020jqw} and the primordial non-Gaussianity \citep{Reischke:2020cgd} . The advantage of this method is that we only need the normalized redshift distributions of FRBs catalog, instead of the precise redshift information for each FRB. In this paper, we study the auto-correlation of the fluctuations of DM from mock FRB samples span $z=0$ to $z=3$. Together with the Planck 2018 temperature and polarization measurements, we study the constraints on the evolution of the baryon mass fraction in the IGM using both the parametric and the non-parametric methods. The rest of the paper is organized as follows. In Sec. \ref{sec:mt} we introduce the clustering properties of the observed DM and the theoretical model of the auto-correlation analysis. In Sec. \ref{sec:data} we explicit the fitting analysis and the data used in our calculations. The constraint results using different numerical methods are shown in Sec. \ref{sec:result}. Finally, we present conclusions and discussions in Sec. \ref{sec:con}.

\section{Methodology}
\label{sec:mt}
The observed dispersion measure $\DM_{\obs}$ is defined as the integral of free electrons number density  along the line of sight, which consists of the contributions from the intergalactic medium, $\DM_{\IGM}$; the FRB host galaxy, $\DM_{\rm host}$ and the Milky Way, $\DM_{\MW}$. $\DM_{\IGM}$ from a fixed source redshift $z_{\rm s}$ and angular position $\vec{\theta}$ is given by:
\be
\label{eq:DM}
\mathrm{DM}_{\mathrm{IGM}}\left(\vec{\theta}, z_{\rm s}\right)=\int_{0}^{z_{\rm s}} \frac{\mathrm{d} z}{H(z)} \frac{n_{\e}(\vec{\theta}, z)f_{\IGM}(z)}{(1+z)^{2}}~,
\ee
where $n_{\e}(\vec{\theta}, z)$ represents the number density of free electrons which can be expressed as
\be
n_{\e}(\vec{\theta}, z)=\frac{ \rho_{\rm b}(\vec{\theta}, z)}{m_{\rm p}}\left(1-\frac{1}{2}Y\right)~~~({\rm for}~z<3)~,
\ee
where $Y\simeq0.24$ is the mass fraction of helium, $\rho_{\rm b}(\vec{\theta}, z)$ is the baryon mass density, $m_{\rm p}$ is the proton mass, and we have assumed both hydrogen and helium are fully ionized at $z<3$ \citep{Becker:2010cu}.

In order to extract information from $\DM_{\IGM}$, we also need to determine $\DM_{\rm host}$ and $\DM_{\MW}$. For a well-localized FRB, $\DM_{\MW}$ can be determined by Galactic pulsar observations \citep{Taylor:1993my}, but deriving $\DM_{\rm host}$ is quite difficult due to the dependences on the type of the host galaxy, the relative orientations  and the near-source plasma, which are poorly known. As a phenomenological attempt, we model $\DM_{\rm host}$ as a function of redshift by assuming the rest-frame measurements accommodate the evolution of star formation history \citep{Luo:2018tiy, Li:2019klc, Wei:2019uhh}, i.e.,
\be
\mathrm{DM}_{\text {host}}(z)=\mathrm{DM}_{\text {host}, 0} \sqrt{\frac{\mathrm{SFR}(z)}{\mathrm{SFR}(0)}}~,
\ee
where we use the present value $\mathrm{DM}_{\text {host}, 0} = 100 ~\rm pc/cm^3$  and ${\rm SFR}(z)=\frac{0.0156+0.118z}{1+(z/3.23)^{4.66}}~{\rm M_{\odot}/yr}$ adopted in \citet{Hopkins:2006bw} and \citet{Wei:2019uhh}
{as our fiducial model.}

In order to preform the 2D spherical projection, we only need the normalized number distribution of FRBs catalog $n(z)$ which can be roughly derived from the $\DM_{\obs}$ \citep{Luo:2018tiy, Zhang:2018fxt, Batten:2020pdh, Takahashi:2021}, instead of the precise redshift information of each sample. Then, $\DM_{\IGM}$, $\DM_{\rm host}$ and $\DM_{\MW}$ for an angular position $\vec{\theta}$ can be written as,
\ba
\label{eq:wd1}
{ \DM}_{{\IGM}}(\vec{\theta}) &=& \int_0^{\infty} {{\rm d} z} W_{{\DM},{\IGM}}(z) \left[1+\delta_{\rm b}(\vec{\theta}, z)\right]~, \\
\label{eq:wd2}
{\mathrm{DM}}_{\mathrm{host}}(\vec{\theta})&=&\int_{0}^{\infty} \mathrm{d} z W_{\mathrm{DM}, \mathrm{host}}(z)\left[1+\delta_{\rm s}(\vec{\theta}, z)\right]~,
\\
{ \DM}_{{\MW}}(\vec{\theta}) &=& \DM_\MW(\vec{\theta})
\ea
where $\delta_{\rm b}$ and $\delta_{\rm s}$ are the baryon density perturbation and FRB number density perturbation, respectively.
{Recent work \citep{Takahashi:2021} shows $\delta_{\rm b}$ agrees with $\delta_{\rm m}$ at large scales ($k<1~h\rm Mpc^{-1}$) but is strongly suppressed at small scales. In our analysis, we adopt the fitting baryon bias factor   $b_{\mathrm{b}}(k,z)\equiv {\delta_{\rm b}}/{\delta_{\rm m}}$ obtained by \citet{Takahashi:2021}.
As for FRB number density perturbation, we assume FRBs form in dark matter halos, so we can express the FRB bias $b_{\rm FRB}(M_{\rm FRB,h}, z) \equiv \delta_{\rm s}/\delta_{\rm m}$ using the fitting formula from \citet{Tinker:2010my}, where the halo mass is set to $M_{\rm FRB,h}=10^{13}h^{-1}M_{\odot}$ as our fiducial model.}
The window functions of Eq.(\ref{eq:wd1}) and Eq.(\ref{eq:wd2}) are
\ba
\label{eq:IGM}
W_{{\DM},{\IGM}}(z)&=&\left(1-\frac{1}{2}Y\right)f_{\IGM}(z) \frac{ \bar\rho_{\rm b,0}}{m_{\rm p}}\frac{{(1+z)}}{H(z)} \int_z^{\infty} n(z) {\rm d}z~, \\
W_{\mathrm{DM}, \text {host}}(z)&=&\frac{\mathrm{DM}_{\text {host}}(z)}{(1+z)} n(z)~,
\ea
where $\bar\rho_{\rm b,0}$ is the average baryon mass density at present time, and we have converted the $\DM_{\rm host}$ from the rest-frame observer to that of the Earth observer by a factor of $1/(1+z)$ \citep{Ioka:2003fr}.

Finally, the auto-correlation power spectrum of the fluctuations of ${\DM}_{\obs}$: $\delta {\DM}_{\obs}(\vec{\theta})\equiv {\DM}_{\obs}(\vec{\theta})-{\overline \DM}_{\obs}$ is given by
\begin{equation}\begin{aligned}
C_{\ell}^{\mathrm{DM}}=& C_{\ell}^{\mathrm{IGM}, \mathrm{IGM}}+C_{\ell}^{\mathrm{host}, \mathrm{host}}+C_{\ell}^{\mathrm{MW}, \mathrm{MW}}+\\
& C_{\ell}^{\mathrm{IGM}, \mathrm{host}}+C_{\ell}^{\mathrm{IGM}, \mathrm{MW}}+C_{\ell}^{\mathrm{host}, \mathrm{MW}}~.
\end{aligned}\end{equation}
Usually, we do not consider the contributions from $\delta {\DM}_{\MW}(\vec{\theta})$ and its correlations with $\delta {\DM}_{\IGM}(\vec{\theta})$ and $\delta {\DM}_{\rm host}(\vec{\theta})$. Therefore, we have three terms left: $C^{\IGM,\IGM}_{\ell}, C^{{\rm IGM},{\rm host}}_{\ell}$ and $C^{{\rm host},{\rm host}}_{\ell}$. Based on the Limber approximation \citep{limber1953analysis}, we have
\be
\label{eq:ps}
C^{\mathrm{IGM},\mathrm{IGM}}_{\ell}=\int \mathrm{d} z W_{\mathrm{DM}, \mathrm{IGM}}^{2}(z) \frac{H(z)}{\chi^{2}(z)} b^2_{\rm b}P_{\rm m}\left( \frac{\ell+1/2}{\chi(z)},z \right)~,
\ee
\be
\begin{aligned}
C_{\ell}^{\text {IGM,host}}=& 2\int \mathrm{d} z \, W_{\text {DM}, \text {IGM}}(z) W_{\text {DM}, \text {host}}(z) \frac{H(z)}{\chi^{2}(z)} \\
& \times b_{\text {FRB}}b_{\rm b} P_{\rm m}\left(\frac{\ell+1 / 2}{\chi(z)}, z\right)~,
\end{aligned}
\ee
\be
C^{\text {host,host}}_{\ell}=\int \mathrm{d} z W_{\mathrm{DM}, \text {host}}^{2}(z) \frac{H(z)}{\chi^{2}(z)} b_{\rm FRB}^2P_{\rm m}\left( \frac{\ell+1/2}{\chi(z)},z \right)~,
\ee
where $\chi(z)$ is the comoving distance and $P_{\rm m}$ is the matter power spectrum.

\section{Data and Likelihood}
\label{sec:data}

In our analysis, we includes the measurements of CMB temperature and polarization anisotropy from the Planck 2018 legacy data release \citep{Aghanim:2018eyx}, which are used to constrain the $\Lambda$CDM parameters. We use the combination of the \texttt{Plik} likelihood using $TT$, $TE$ and $EE$ spectra at $\ell\ge30$, the low-$\ell$ ($\ell=2\sim29$) temperature \texttt{Commander} likelihood and the \texttt{SimAll} $EE$ likelihood, which is labeled as TT,TE,EE+lowE in \citet{Aghanim:2018eyx}.

Then, we discuss the mock angular power spectrum used in this paper. We do not need the precise redshift information of each FRB which is hard to obtain. Instead, we can use the observed $\DM$s to estimate the redshift distribution $n(z)$. Here, we assume the redshift distribution of FRBs in the redshift range $0<z<3$ is $n(z)\propto z^{k}{\rm e}^{-z/\lambda}$ \citep{Zhou:2014yta}, {and we set $k=1$ and $\lambda=1$ as our fiducial model}.

We also need to consider the noise spectrum for the observed $\delta {\DM}_{\obs}(\vec \theta)$, which can be decomposed as
\be
\label{eq:nps}
N^{\DM}_{\ell} = \sqrt{\frac{2}{(2\ell+1)f_{\rm sky}}} \left[C^{\DM}_{\ell}+N^{\rm host}_{\ell}\right]~,
\ee
where $N^{\rm host}_{\ell}$ is the noise induced by the intrinsic scatter of DM around host galaxies \citep{Shirasaki:2017otr, Reischke:2020cgd, Takahashi:2021}: $N^{\rm host}_{\ell} = {4\pi}f_{\rm sky}\sigma^2_{\rm host}/{\mathcal{N}}$. We set the intrinsic scatter of DM around host galaxies $\sigma_{\rm host}=30 \rm~pc/cm^{3}$ (redshift independent, i.e., $30(1+z) \rm~pc/cm^{3}$ in the rest-frame), the sky fraction {$f_{\rm sky}=0.8$} and we use $\mathcal{N}=10^4$ FRBs span $z=0$ to $z=3$. The mock DM angular power spectrum at $\ell$ can be easily obtained by Gaussian sampling with the mean value $\mu=C_{\ell}^{\DM}$ and the standard deviation $\sigma=N_{\ell}^{\DM}$.

Finally we can obtain the $\chi^2$ function, where we have assumed the different scales are independent with each other,
\be
\chi^2=\left(\hat{C}_{\ell}^{\DM}-C_{\ell}^{\DM}\right) \Gamma^{-1}_{\ell,\ell'}\left(\hat{C}_{\ell'}^{\DM}-C_{\ell'}^{\DM}\right)^{\rm T}~,
\ee
where $C_\ell^{\DM}$ refer to the theoretical model and $\hat{C}_\ell^{\DM}$ is our mock spectrum. { $\Gamma_{\ell,\ell'} = \delta_{\ell, \ell'} (N_{\ell}^{\rm DM})^2$ is the diagonal covariance matrix}. In our analysis, we set $\ell_{\max}=500$ due to the resolution limitation of the observed FRBs.

{
With these data likelihoods, we preform a global fitting analysis using the \textsc{CosmoMC} package \citep{Lewis:2002ah}, a Markov Chain Monte Carlo (MCMC) code with a purely adiabatic initial conditions and a $\Lambda$CDM universe. The parameterization used in our analysis is thus: $P\equiv \{\Omega_{\rm b} h^{2}, \Omega_{\rm c} h^{2}, \Theta_{\rm s}, n_{\rm s}, A_{\rm s}, P_{\rm IGM}, M_{\rm FRB,h}, \rm {DM_{host,0}}, \Delta z\}$, where $\Omega_{\rm b} h^{2}$ and $\Omega_{\rm c} h^{2}$ are the baryon and cold dark matter physical density, $\Theta_{\rm s}$ is the angular size of the sound horizon at decoupling, $n_{\rm s}$ and $A_{\rm s}$ are the spectral index and the primordial power spectrum, and $P_{\rm IGM}$ are the parameters which describe the evolution of $f_{\IGM}(z)$.}
{
We also include three additional nuisance parameters $\{M_{\rm FRB,h}, \rm {DM_{host,0}}, \Delta z \}$ to make a more robust conclusion. $M_{\rm FRB,h}$ is used to account for the FRB bias uncertainty. We set this parameter vary between $[10^{12.5}, 10^{13.2}]~h^{-1}M_{\odot}$, and the spread of FRB bias $\Delta b_{\rm FRB}(z)/b_{\rm FRB}(z)$ is about 30\%$\sim$60\%. Since $b_{\rm FRB}$ can be obtained by the auto-correlation of FRB density field and cross-correlation between FRB and galaxy density field (e.g. \citep{Shirasaki:2017otr}), this parameter space is sufficient to include the effect of $b_{\rm FRB}$ with future surveys. We also vary $\rm {DM_{host,0}}$ between [70,130] $\rm pc/cm^3$ in our MCMC analysis. With more well-localized FRBs in the future, the origin and the host environment of FRB will be better known, so the parameter space can be also sufficient in this work.
}
{
The last nuisance parameter $\Delta z$ is used to describe the uncertainty of $n(z)$. Recent simulations (e.g. \citep{Batten:2020pdh, Takahashi:2021}) show the probability distribution of DM at a given redshift is skewed and the inferred $z_s$ is higher than the analytical mean for a given DM, thus induce a bias on $n(z)$. In our analysis, we take this bias into account by shifting the redshift distribution from $n(z)$ to $n(z+\Delta z)$, and we have checked that using $\Delta z = 0.1$ can eliminate this bias well. We set $\Delta z$ vary between [-0.1, 0.1] to include this effect. What is more, the redshift distribution of FRBs also can be estimated from an empirical relation in the era of Square Kilometre Array (SKA) \citep{Hashimoto2021}, which can give us a complementary check.
}

\section{Constrain results}
\label{sec:result}
\subsection{Parametric method}
\begin{figure}
	\centering
    \includegraphics[width=1\linewidth]{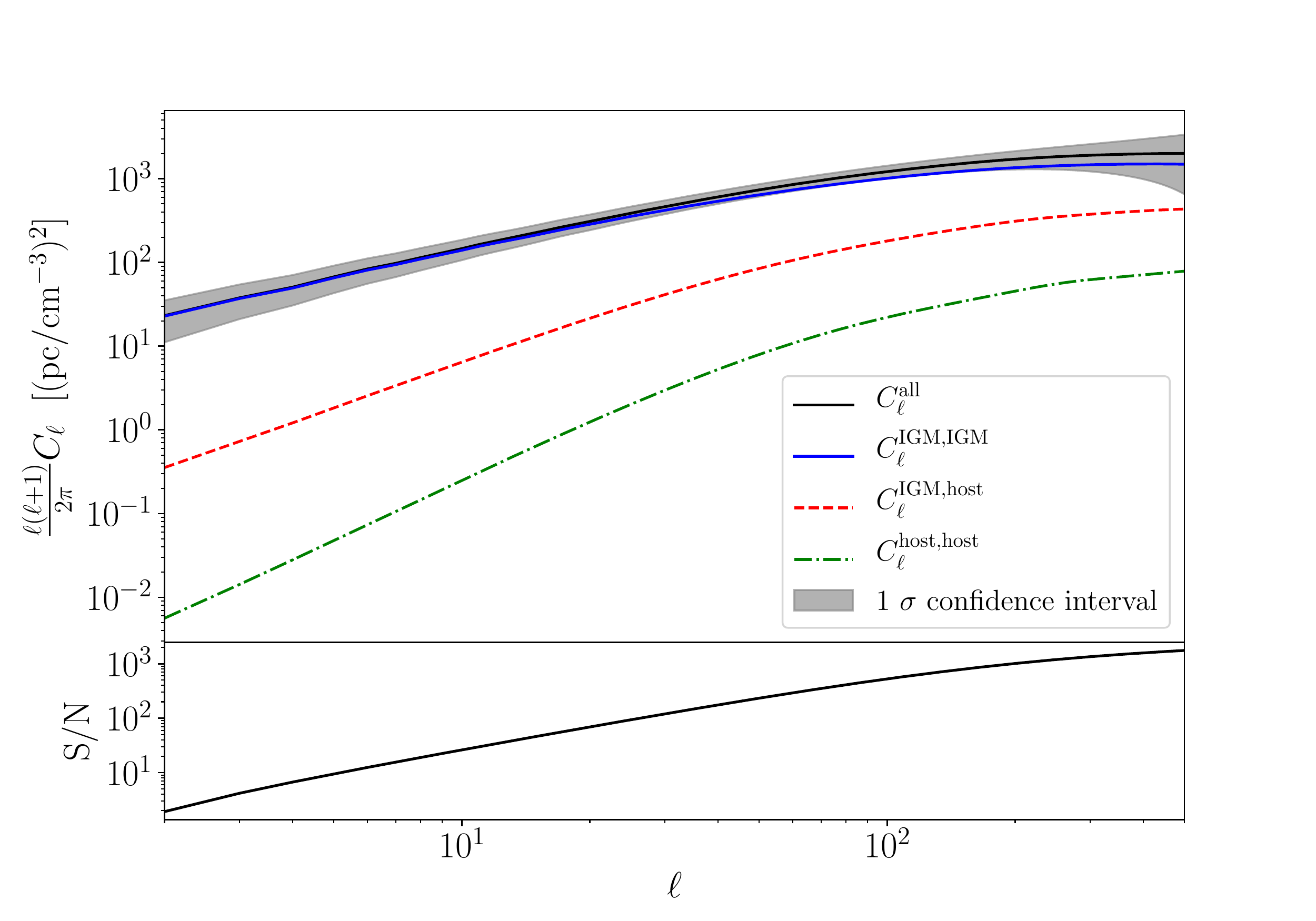}
	\caption{Top: The angular power spectra of DM from the auto-correlation of the IGM component (blue solid line), the cross-correlation of the IGM and the host galaxy component (red dashed line), and the auto-correlation of the host galaxy component (green dotted-dashed line). {We also plot their summation (black solid line) and its $1\sigma$  confidence interval. Bottom: Sinal to noise ratio S/N as function of $\ell_{\max}$.}}
	\label{fig:cl}
\end{figure}

First, we consider a parametric method by expanding $f_{\IGM}(z)$ into Taylor series up to second order \citep{Li:2019klc, Wei:2019uhh}:
\be
\label{eq:pm}
f_{\IGM}(z)=f_{\IGM,0}+\alpha\frac{z}{1+z}~,
\ee
where $f_{\IGM,0}$ and $\alpha$ are two free parameters. To generate the mock spectrum, we fix the fiducial values $f_{\IGM,0}=0.75$, $\alpha=0.25$ since $f_{\IGM}(z)$ is slowly growing with redshift, and other cosmological parameters are fixed to be the best fit values from Planck 2018 results using TT,TE,EE+lowE. In Fig. \ref{fig:cl} we show the auto-correlation power spectra from different components and we find the contribution from the IGM component dominates the signal.
{We also show the $1\sigma$  confidence interval and sinal to noise ratio ${\rm{S/N}} \equiv \sum_{\ell=2}^{\ell_{\max}}C_\ell^{\rm DM}/N_\ell^{\rm DM}$ as function of $\ell_{\max}$ in Fig. \ref{fig:cl}}

\begin{figure}
	\centering
    \includegraphics[width=1\linewidth]{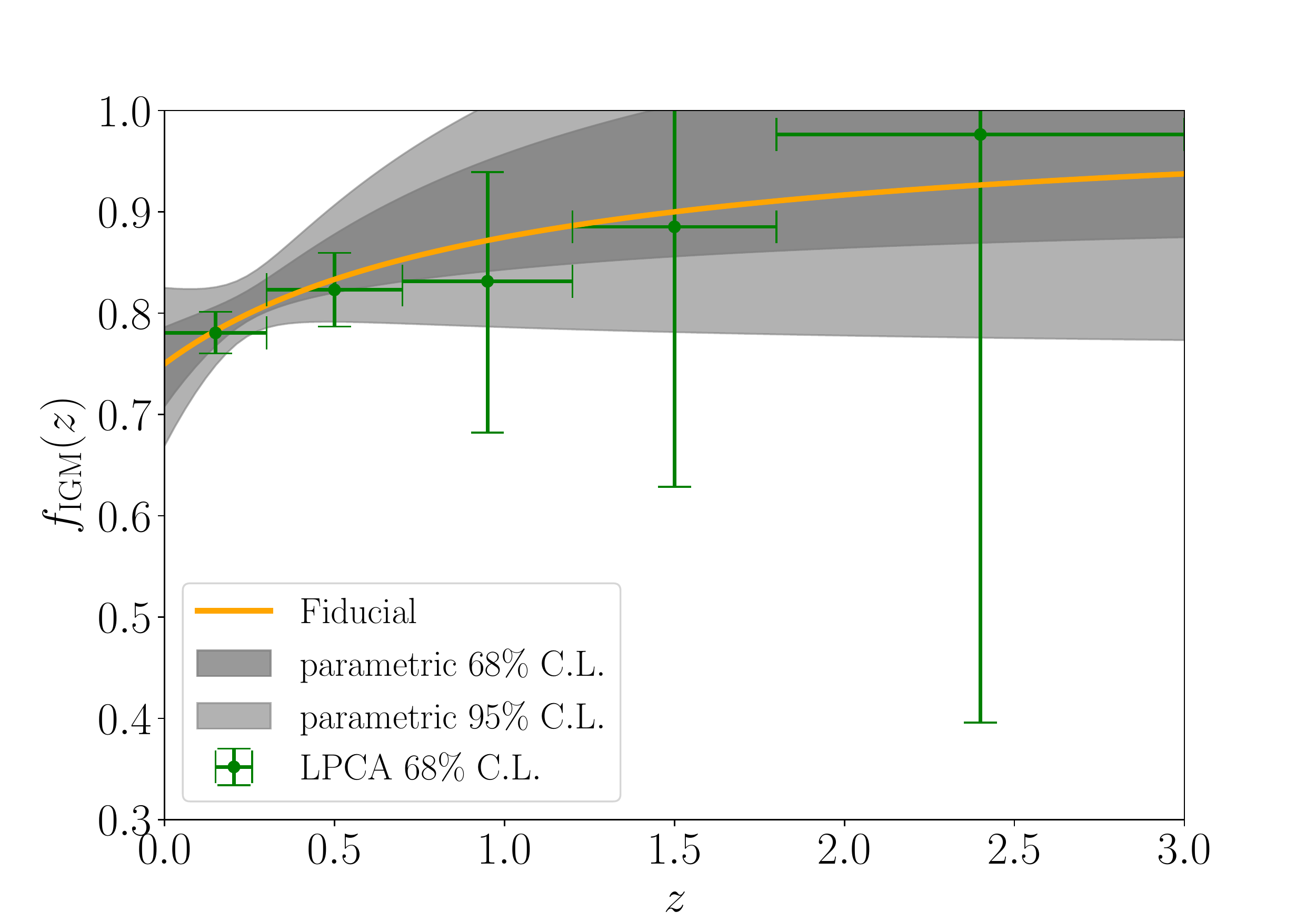}
	\caption{The reconstructed $f_{\IGM}(z)$ with different methods: the orange line is the fiducial model, the grey areas are the $68\%$ and $95\%$ confidence regions from the parametric method, the green error-bars are the $68\%$ C.L. obtained from the LPCA method.}
	\label{fig:env2p}
\end{figure}

Together with the Planck 2018 temperature and polarization measurements, our mock DM angular power spectrum could give tight constraints on the parameters: {$f_{\rm IGM,0} = 0.747\pm 0.039$ and $\alpha = 0.306 \pm 0.182$} at 68\% confidence level, which are comparable with the previous works \citep{Li:2019klc, Wei:2019uhh}. In order to present the constraints on the $f_{\IGM}$ at different redshift, we use
\be
\begin{aligned}
\sigma^2[f_{\IGM}(z)] &\equiv \sigma^2(f_{\IGM,0})+\sigma^2(\alpha)\left(\frac{z}{1+z}\right)^2 \\
&+2{\rm COV}(f_{\IGM,0},\alpha)\left(\frac{z}{1+z}\right)
\end{aligned}
\ee
to reconstruct the redshift evolution of $f_{\IGM}(z)$ in Fig. \ref{fig:env2p}, where ${\rm COV}(f_{\IGM,0},\alpha)$ is the covariance  between  $f_{\IGM,0}$ and $\alpha$. The orange line denotes the evolution of the fiducial model, and the grey areas are the 68\% and 95\% confidence intervals using this parametric method. Finally we can quantify the reconstructed uncertainty by calculating  {$\sigma(f_{\IGM})\equiv \left\langle \sigma [f_{\IGM}(z)] /f_{\IGM}(z) \right\rangle$ which is about 7.2\%}. This result with $10^4$ FRBs is very promising, which do not need the precise redshift information of each sample.

\subsection{Non-parametric method}

Although we already have a sufficient constraint on $f_{\IGM}(z)$, we have to assume the evolution form first, which may leads to some biases to getting the true value. In order to conduct a parameterization-independent analysis, the simplest way is dividing the redshift range into several bins and assume $f_{\IGM}$ as a constant in each redshift bin. However, due to the strong degeneration between different bins, we can hardly obtain the proper constraints. To solve this problem, we adopt the so-called  ``local principle component analysis" (LPCA) method which has been used in the analysis of uncorrelated galaxies power spectrum \citep{Hamilton:1999uw} and equation of state of dark energy \citep{Huterer:2004ch, Zhao:2007ew, Zheng:2014ara, Dai:2018zwv}.

In our analysis, we divide the redshift range into 5 bins with $0\le z_1<0.3$, $0.3\le z_2<0.7$, $0.7\le z_3<1.2$, $1.2\le z_4<1.8$, $1.8\le z_5<3$ and assume the corresponding parameter $f_{\IGM}(z_i)$ are constants. From MCMC calculation, we can compute the covariance matrix $C$ of $f_{\IGM}(z_i)$ by marginalizing over the other parameters, and get the Fisher matrix $F=C^{-1}$. In order to get uncorrelated $f_{\IGM}(z_i)$, we should rotate the original vectors into a basis where the covariance matrix is diagonal. To achieve this, we can diagonalize the Fisher matrix using an orthogonal matrix $W$, $F=W^{\rm T}\Lambda W$. Furthermore, we can define $\hat W$ by absorbing the diagonal matrix $\Lambda^{1/2}$ into $W$, so we have $\hat{W}^{\rm T}\hat W=F$. After normalizing $\hat W$, we find the weights (rows of $\hat W$) are almost everywhere positive and the $i$th element of the $i$th weight has the maximum value. Then the new parameters $\hat f_{\IGM}(z_i)$, defined as $\hat f_{\IGM}(z_i) = \hat W f_{\IGM}(z_i)$, are uncorrelated, since they have the diagonal covariance matrix.

Using the LPCA method, we can rotate the constraints on $f_{\IGM}(z_i)$ obtained from the MCMC into the uncorrelated constraints on $\hat f_{\IGM}(z_i)$, whose marginalized $1\sigma$ constraint results are: {$\hat f_{\IGM}(z_1)=0.781\pm0.021, \hat f_{\IGM}(z_2)=0.823\pm0.039, \hat f_{\IGM}(z_3)=0.831\pm0.128, \hat f_{\IGM}(z_4)=0.885\pm0.265, \hat f_{\IGM}(z_5)=0.976\pm0.672$}, which are shown in Fig. \ref{fig:env2p} (green error bars). We find the results are consistent with the fiducial model and the constraints using the parametric method, but have larger error-bars increasing with the redshift. Based on Eq.(\ref{eq:wd1}) and Eq.(\ref{eq:IGM}), we know that, since the windows function becomes smaller as the redshift increases, the auto-correlation power spectrum is not very sensitive to $f_{\IGM}$ at higher redshifts. Consequently, the obtained constraints on $f_{\IGM}$ will also become weaker at high redshifts. The main contribution of the constraining power comes from the first three bins. We also compute the mean relative error of these five $\hat f_{\IGM}$ at different redshift bins, and obtain that {$\sigma(\hat f_{\IGM})=\left\langle \sigma[ \hat f_{\IGM}(z_i)]/ \hat f_{\IGM}(z_i) \right\rangle\simeq24.2\%$}, which is {three} times larger than that using the parametric method.

\section{Conclusions and discussions }
\label{sec:con}
In this paper, we apply the auto-correlation power spectrum of $\delta\DM$ from the mock DM angular power spectrum to constrain the baryon mass fraction in the IGM. The main advantage of this method is that the precise redshift measurements are not necessary. We only need a rough redshift distribution from observations. Using the auto-correlation power spectrum from $10^4$ FRBs with the host DM scatter of $30(1+z)~\rm pc/cm^3$ spanning {80\%} of the sky and redshift range $z=0-3$, together with the Planck 2018 temperature and polarization measurements, we can obtain a very tight constraint on the evolution of $f_{\IGM}(z)$ if we use the simple Taylor expansion up to second order, and the reconstructed mean relative standard deviation is {$\sigma(f_{\IGM})\simeq7.2\%$}. Furthermore, we adopt the LPCA methods to constrain the evolution of $f_{\IGM}$. We obtain very weak constraints on $f_{\IGM}(z_i)$ at higher redshifts, which leads to a larger uncertainty {$\sigma(f_{\IGM})\simeq24.2\%$}.

{
We must mention that the intrinsic scatter of DM around host galaxies $\sigma_{\rm host}$ is still uncertain. To demonstrate the robustness of our analysis, we also use $\sigma_{\rm host} = 100~\rm pc/cm^3$ to constrain the parametric model (Eq.(\ref{eq:pm})). The constraints are: $f_{\rm IGM,0} = 0.752\pm 0.092$ and $\alpha = 0.283 \pm 0.394$, which are twice times larger than the results when $\sigma_{\rm host} =30 ~\rm pc/cm^3$. Recent works (e.g. \citep{Hashimoto:2020dud}) show FRBs will be detected with SKA at a rate of $\sim 10^3-10^4~\rm (sky^{-1} day^{-1})$. Since shot noise is proportion to $\sigma^2_{\rm host}/N$, $10^5$ FRBs with the intrinsic scatter $\sigma_{\rm host} =100~ \rm pc/cm^3$ may have comparable results obtained in our analysis, and it is accessible with the future SKA survey.
}

{
Finally, we discuss if the expected constraints on $f_{\rm IGM}$ can exclude some models. The ideal case is we can directly select models with the non-parametric method. In Fig. \ref{fig:model} we plot five mean DM-$z$ relations. Results from \citet{Ioka:2003fr} and \citet{Zhang:2018fxt} are analytical formulations. \citet{Ioka:2003fr} assumes the Universe is homogeneously filled with ionised hydrogen alone, and \citet{Zhang:2018fxt} includes helium reionization and uses $f_{\rm IGM} = 0.85$ to exclude baryons locked inside galaxies. These two models can be seen as the upper-limit and lower-limit to the slope of the DM-$z$ relation \citep{Batten:2020pdh}. On the other hand, \citet{McQuinn:2013tmc, Dolag:2014bca, Batten:2020pdh} both use simulations to estimate the DM-$z$ relations. In the bottom panel of Fig. \ref{fig:model} we plot the relative differences between \citet{Batten:2020pdh}'s result and the others results $\rm |DM_{others} - DM_{Batten}|/DM_{Batten}$. To check the ability to select models, we need compare these relative differences with relative confidence interval of DM$(z)$ obtained by LPCA method. In our analysis, we calculate $\DM(z)$ for each step in the MCMC calculation and estimate  $\sigma\DM(z)/\DM(z)$ at each redshift.   The result is also shown in the bottom panel of Fig. \ref{fig:model} (gray region). We can find the expected constraints on $f_{\IGM}$ in our work can be used to exclude models, especially at lower redshifts.
\begin{figure}
	\centering
    \includegraphics[width=1\linewidth]{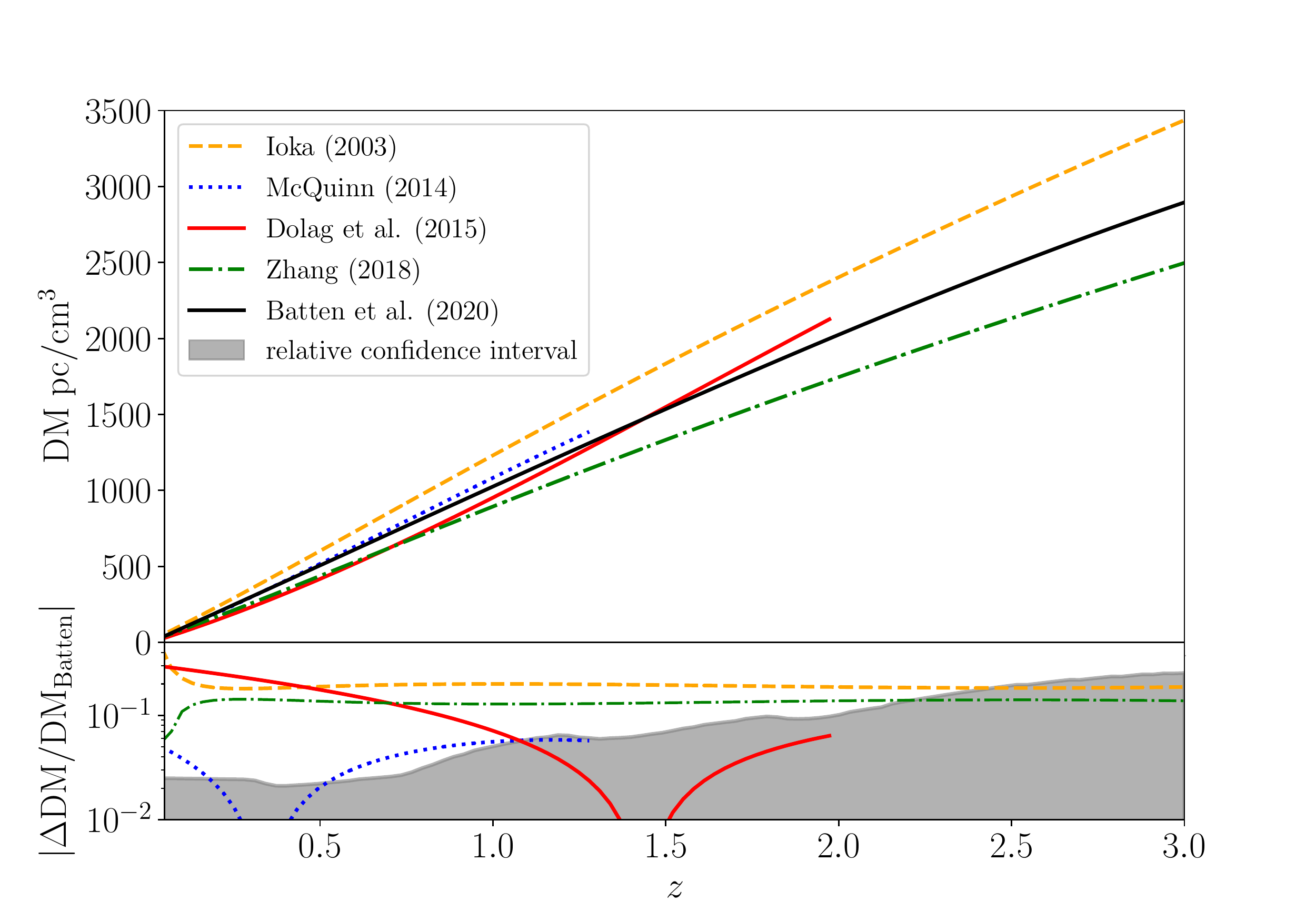}
	\caption{Top: The mean DM-$z$ relations from the past works. Bottom: The relative differences between \citet{Batten:2020pdh}'s result and the others results. The gray region is the relative $1\sigma$ confidence interval of DM$(z)$ in our work using LPCA method.}
	\label{fig:model}
\end{figure}
}

With the forthcoming surveys, this could be a complimentary method to investigate the baryon mass fraction in the IGM and the other cosmology problems, e.g., the equation of state of dark energy, Hubble constant, etc.

\section{DATA AVAILABILITY}
The  measurements of CMB temperature and polarization anisotropy from the Planck 2018 legacy data release are available in \citet{Aghanim:2018eyx} and can be downloaded from \url{http://pla.esac.esa.int/pla/index.html#home}. The mock DM angular power spectrum will be shared on reasonable request to the corresponding author.

\section*{Acknowledgements}
We thank Z.-X. Li and H. Gao for useful discussions. This work is supported by the National Science Foundation
of China under grants No. U1931202 and 12021003, and the National Key R\&D Program of China under
grant No. 2017YFA0402600.


\bsp	
\label{lastpage}
\end{document}